\documentclass[a4paper,11pt]{article}
\usepackage{jinstpub} 
\usepackage{xspace}

\newcommand{\fig}[1]{figure~\ref{fig:#1}}

\newcommand{\sect}[1]{section~\ref{sec:#1}}

\newcommand{\ivalue}{\mbox{I-value}\xspace}
\newcommand{\ivalues}{\mbox{I-values}\xspace}
\newcommand{\geant}{\textsc{Geant4}\xspace}

\title{Measurement of the mean excitation energy of liquid argon}

\author{M. Strait}
\affiliation{Fermilab, PO Box 500, Batavia IL 60510, USA}

\emailAdd{mstrait@fnal.gov}

\newcommand{\garanswer}{$(187\pm 4)$\,eV\xspace}

\newcommand{\laranswer}{$(197\pm 7)$\,eV\xspace}
\newcommand{\laranswertwodigits}{$(196.8\pm 6.5)$\,eV\xspace}

\newcommand{\larmeas}{$(188\pm8)$\,eV\xspace}
\newcommand{\larmeastwodigits}{$188.2^{+7.5}_{-8.0}$\,eV\xspace}

\newcommand{\recommendation}{$(193\pm5)$\,eV\xspace}

\abstract{The mean excitation energy (I-value) of liquid argon is
a critical input for energy estimation in neutrino oscillation experiments.  It is
measured to be $(188\pm8)$\,eV using the range of 404\,MeV protons from
the Fermilab Linac.  This compares to the author's recent evaluation of
$(197\pm7)$\,eV based on a combination of an oscillator strength distribution
analysis, gaseous argon range measurements, sparse stopping power data on solid
argon, and an extrapolation of data on the effect of phase from other
substances.  Using all sources of information, we recommend a value of
$(193\pm5)$\,eV for liquid argon.}


\arxivnumber{2501.10550}

\begin{document}
\maketitle
\flushbottom

\frenchspacing

\newcommand{\tableorder}
{
  \begin{table}

  \caption{Data taking.  The 31 pulses are shown, giving the local time and 
   copper thickness used for each.}
  \label{tab:order}

  \centering

  \begin{tabular}{c r | c r | c r}

  \hline
  \hline

  Time & Copper & Time & Copper & Time & Copper \\

  \hline

    11:18 & 21.5\,mm &	    11:34 & 21.5\,mm &	    12:52 & 17.5\,mm \\
    11:24 & 11.9\,mm &	    11:35 & 21.5\,mm &	    12:53 & 17.5\,mm \\
    11:25 & 11.9\,mm &	    11:36 & 18.3\,mm &	    12:54 & 14.3\,mm \\
    11:26 & 11.9\,mm &	    11:37 & 18.3\,mm &	    12:55 & 14.3\,mm \\
    11:27 & 8.8\,mm  &	    12:43 & 15.1\,mm &	    12:56 & 14.3\,mm \\
    11:28 & 8.8\,mm  &	    12:47 & 20.7\,mm &	    12:57 & 14.3\,mm \\
    11:29 & 8.8\,mm  &	    12:48 & 17.5\,mm &	    12:58 & 14.3\,mm \\
    11:30 & 15.1\,mm &	    12:49 & 17.5\,mm &	    12:59 & 11.1\,mm \\
    11:31 & 15.1\,mm &	    12:50 & 17.5\,mm &	    13:00 & 11.1\,mm \\
    11:32 & 15.1\,mm &	    12:51 & 17.5\,mm &	    13:01 & 11.1\,mm \\
    11:33 & 21.5\,mm \\
 
  \hline
  \hline

  \end{tabular}

  \end{table}
}

\newcommand{\tablesyst}
{
  \begin{table}

  \caption{Systematic uncertainties.  Each category is explained in detail in the
   listed section.}
  \label{tab:syst}

  \centering

  \begin{tabular}{l c c}

  \hline
  \hline

  Uncertainty & eV & Section\\

  \hline

  Beam energy                  & $\pm 6\phantom{.0}$ & \ref{sec:beam} \\
  Scintillator \& camera       & $^{+3.2}
                                  _{-4\phantom{.0}}$ & \ref{sec:scintillator} \\
  Materials accounting         & $\pm 1.2$ & \ref{sec:matcount} \\  
  Multiple Coulomb scattering  & $\pm 1.0$ & \ref{sec:mcs}\\
  Monte Carlo code consistency & $\pm 0.9$ & \ref{sec:mcconsistency} \\
  Aluminum \& copper \ivalues  & $\pm 0.7$ & \ref{sec:otheri} \\
  Other physics                & $\pm 0.5$ & \ref{sec:otherphysics} \\
  Alignment                    & $\pm 0.4$ & \ref{sec:align} \\

  \hline

  Total systematic uncertainty &  $^{+7\phantom{.0}}_{-8\phantom{.0}}$ \\

  Statistical uncertainty & $\pm 1.3$ \\

  \hline

  Total uncertainty & $\pm8$ \\
 
  \hline
  \hline

  \end{tabular}

  \end{table}
}

\newcommand{\figurechromoxspectrum}
{
  \begin{figure}
    \centering

    \includegraphics[width=0.5\columnwidth]{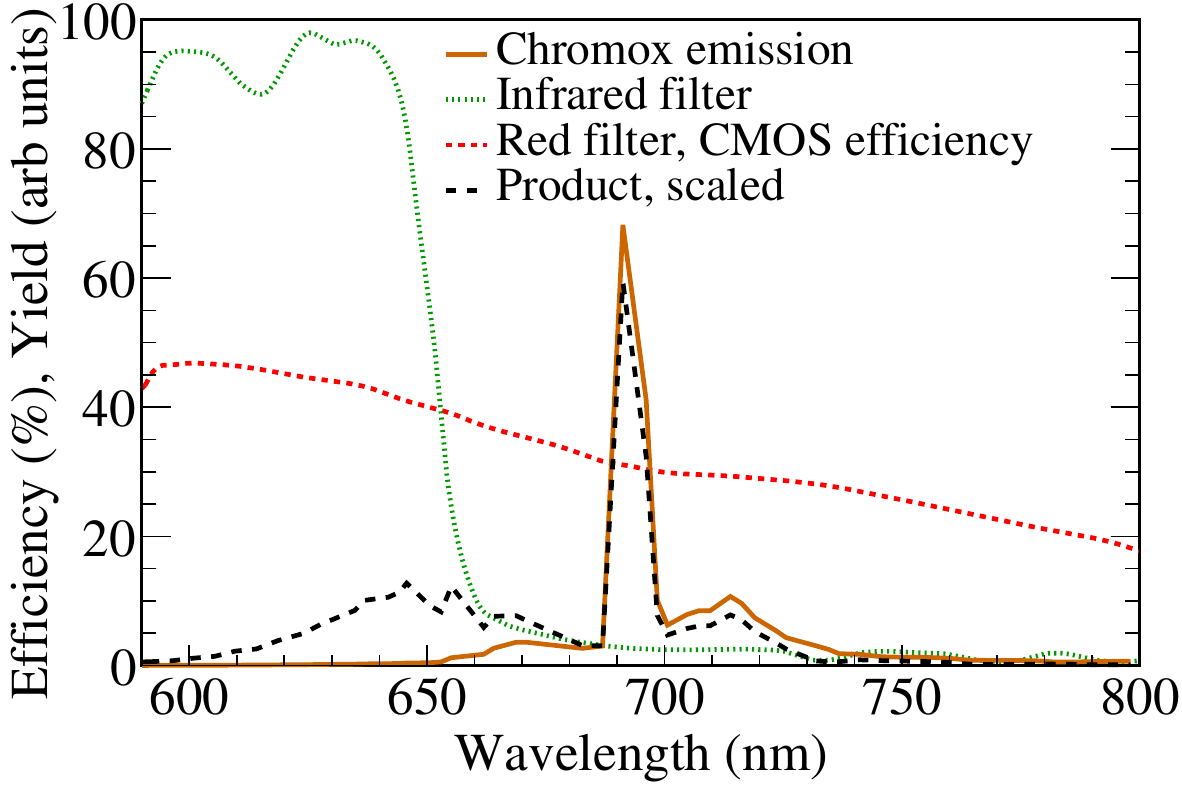}%
    \includegraphics[width=0.5\columnwidth]{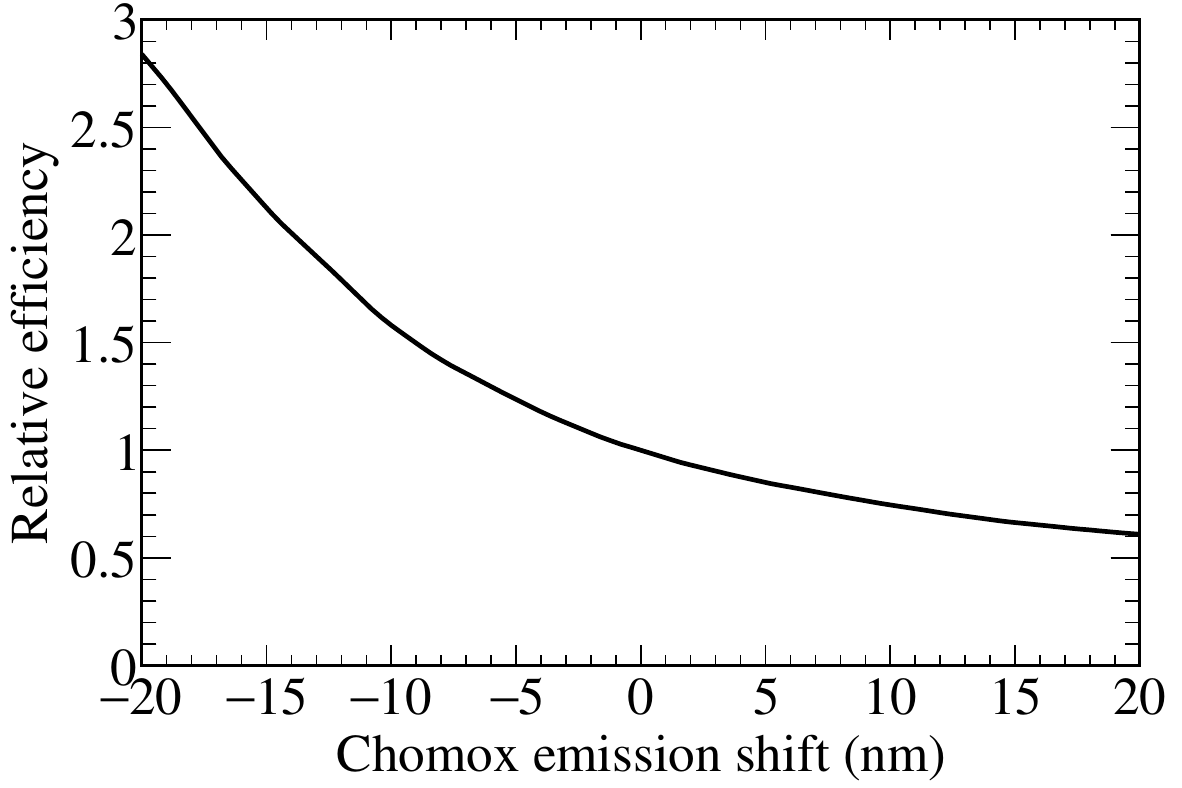}

    \caption{Left: chromox emission spectrum from Ref.~\cite{chromox-xray}, C920 infrared filter transmittance from
             Ref.~\cite{lukselt}, the typical product of red filter
             transmittance and CMOS efficiency for a digital camera, and the overall product of emission and efficiency. Right: resulting overall efficiency, relative to nominal, as a function of
             a uniform shift of the chromox spectrum. }

    \label{fig:shiftchromox}
  \end{figure}
}

\newcommand{\figureshape}
{
  \begin{figure}
    \centering

    \includegraphics[width=0.5\columnwidth]{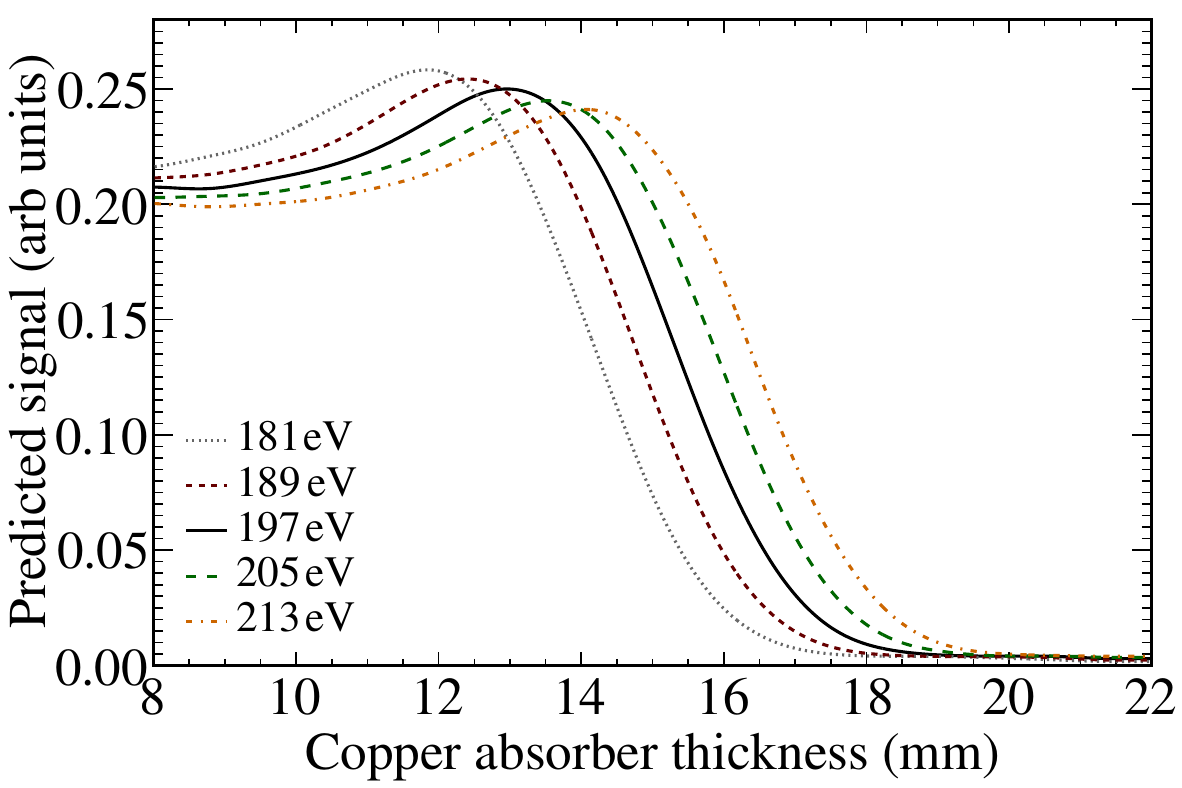}%
    \includegraphics[width=0.5\columnwidth]{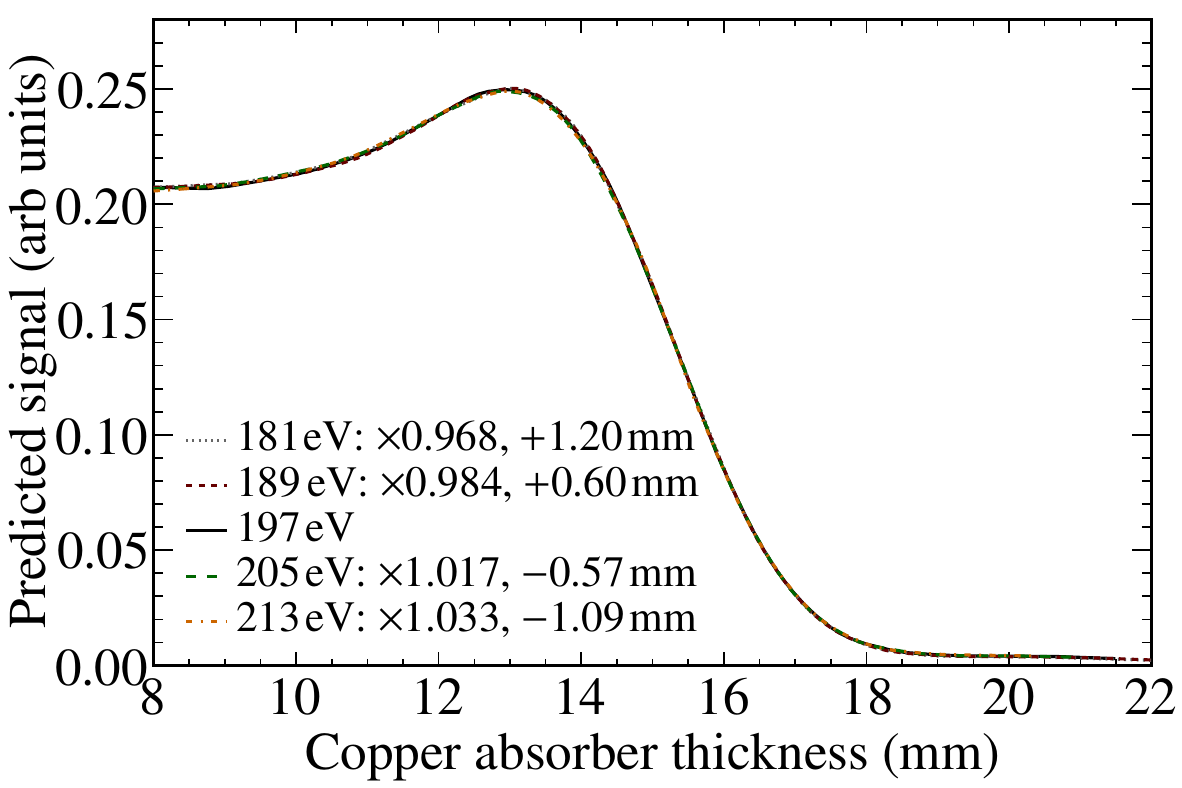}

    \caption{Predicted signal (energy deposition) as a function of \ivalue, showing that only the normalization
             and horizontal position of the curve change with \ivalue, as a good approximation.
              Left: the raw predictions for five \ivalues. Right: the same predictions, scaled
              and translated to match the curve for 197\,eV.}

    \label{fig:shape}
  \end{figure}
}

\newcommand{\figurecook}
{
  \begin{figure}
    \centering

    \includegraphics[width=0.5122\columnwidth]{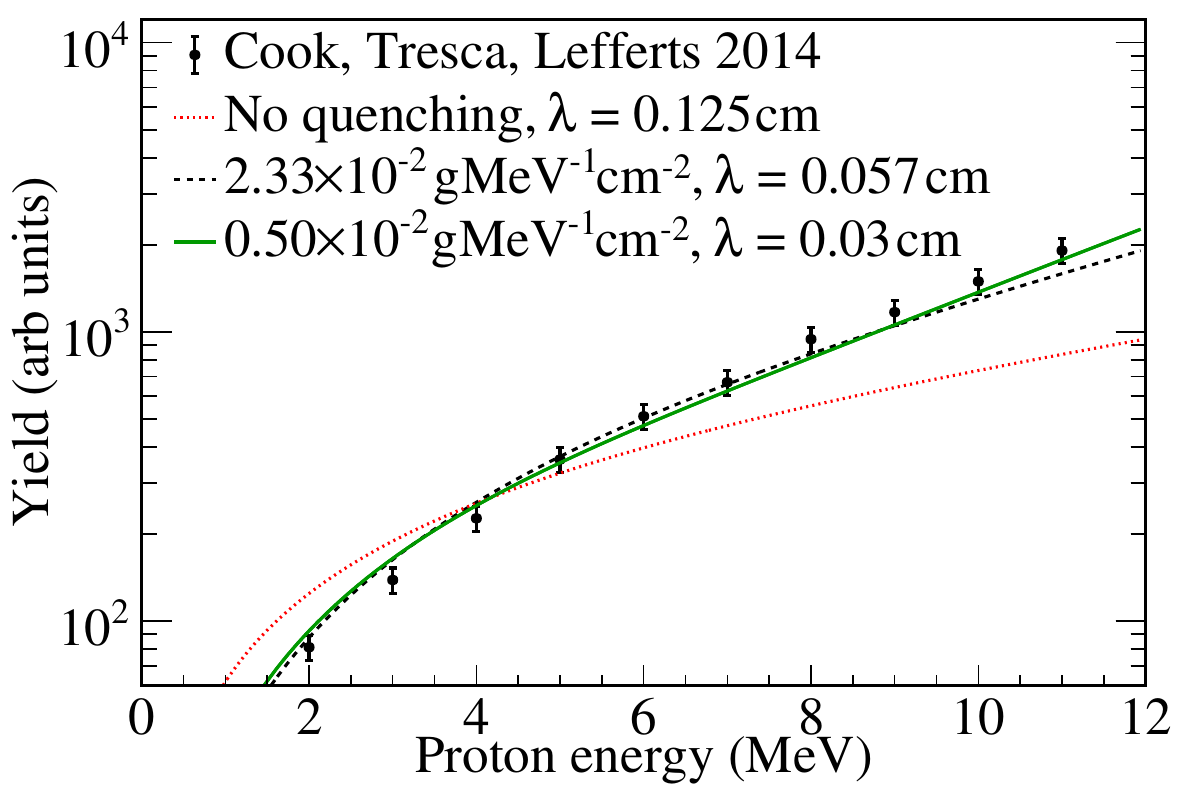}

    \caption{Data from Ref.~\cite{Cook_2014} shown along with our re-simulation of
        their setup for three scenarios: no quenching and the manufacturer's
        quoted attenuation length, Cook's best fit, and our model with
        much less quenching and which
        fits the data better by 0.9 units of $\chi^2$.} 

    \label{fig:cook}
  \end{figure}
}

\newcommand{\figurescheme}
{
  \begin{figure}
    \centering

    \includegraphics[width=1.0\columnwidth]{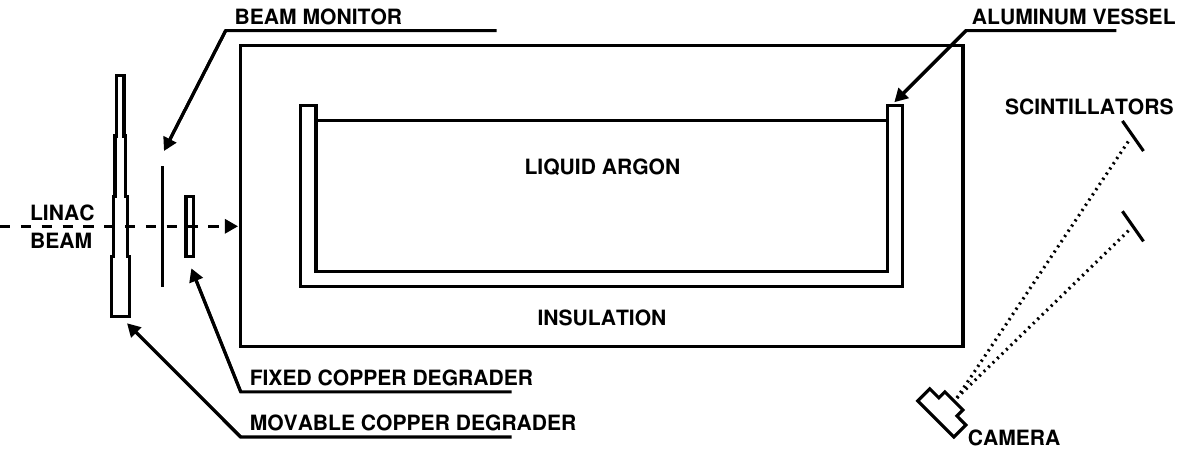}

    ~ 

    \includegraphics[width=4in]{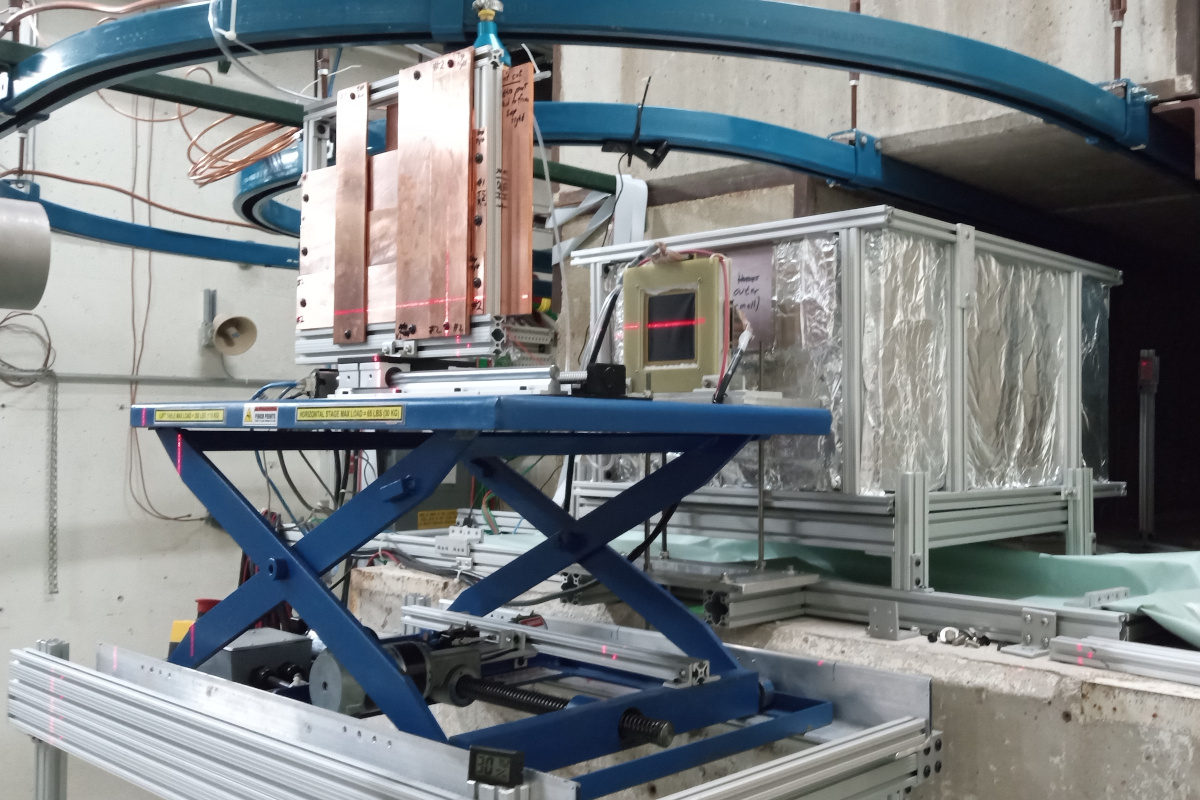}

    \caption{Top: Experimental schematic of experiment carried out at the Fermilab Irradiation Test Area. Not to scale.
             Bottom: photograph of same.}

    \label{fig:scheme}
  \end{figure}
}

\newcommand{\figuremars}
{
  \begin{figure}
    \centering
    \includegraphics[width=0.5\columnwidth]{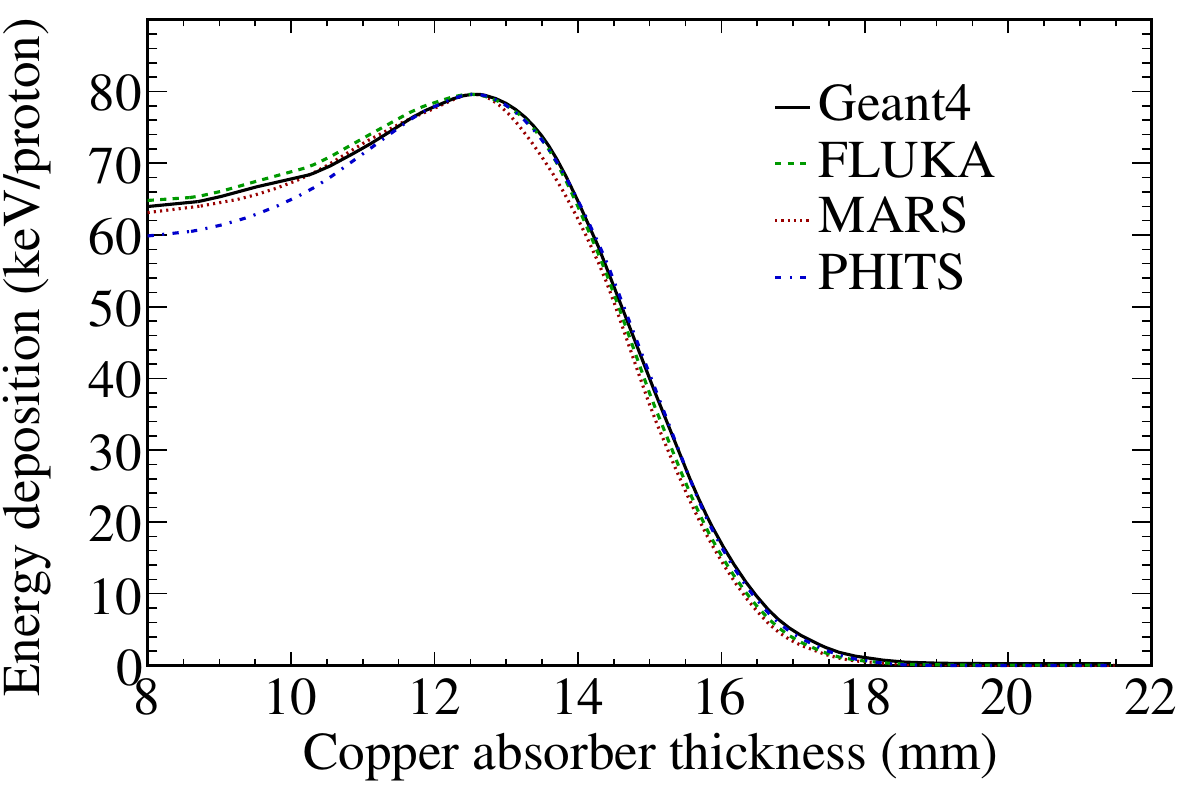}
    \caption{Comparison of several Monte Carlo codes.  The prediction of
             energy deposition in the on-axis scintillator is shown.  The \textsc{mars}, FLUKA and 
             PHITS results have been scaled by $-2.2\%$, $+9.6\%$ and $-0.6\%$, respectively,
             so that the peak matches \geant.}
    \label{fig:mars}
  \end{figure}
}

\newcommand{\figureimages}
{
  \begin{figure}
    \centering

    \includegraphics[width=0.4\columnwidth]{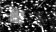} \includegraphics[width=0.4\columnwidth]{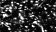}

    \includegraphics[width=0.4\columnwidth]{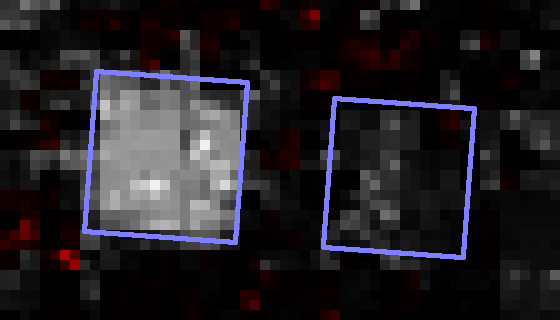}

    \caption{Top left: Example of raw data, red channel.  Top right: same, 44\% of green channel added to 59\% of blue channel.
             Bottom: subtraction of above two images, with outlines showing the positions of the two scintillator squares.
             Pixels with negative values after subtraction are shown in red.  All three images are displayed with increased contrast for clarity.}

    \label{fig:images}
  \end{figure}
}

\newcommand{\figuregeant}
{
  \begin{figure}
    \centering

    \includegraphics[width=\columnwidth]{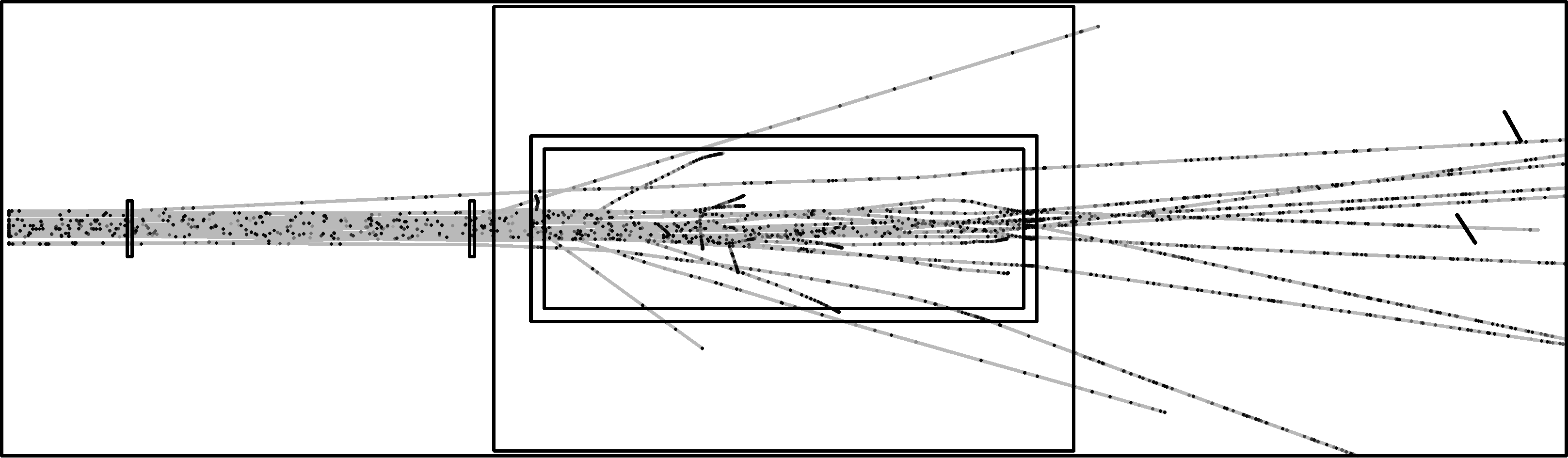}
    \caption{Top view of a \geant simulation of 25 protons in the configuration
       using 11.70\,mm of copper degrader, including the fixed 6.33\,mm
       degrader immediately upstream of the vessel. Gray lines are the
       proton trajectories and the black points on the lines are \geant
       steps.  Secondary particles and the concrete shielding blocks are not shown. 
       The on-axis and off-axis scintillators are visible on
       the right side.}
    \label{fig:geant}

  \end{figure}
}

\newcommand{\figurerealdatafit}
{
  \begin{figure}
    \centering

    \includegraphics[width=0.49\columnwidth]{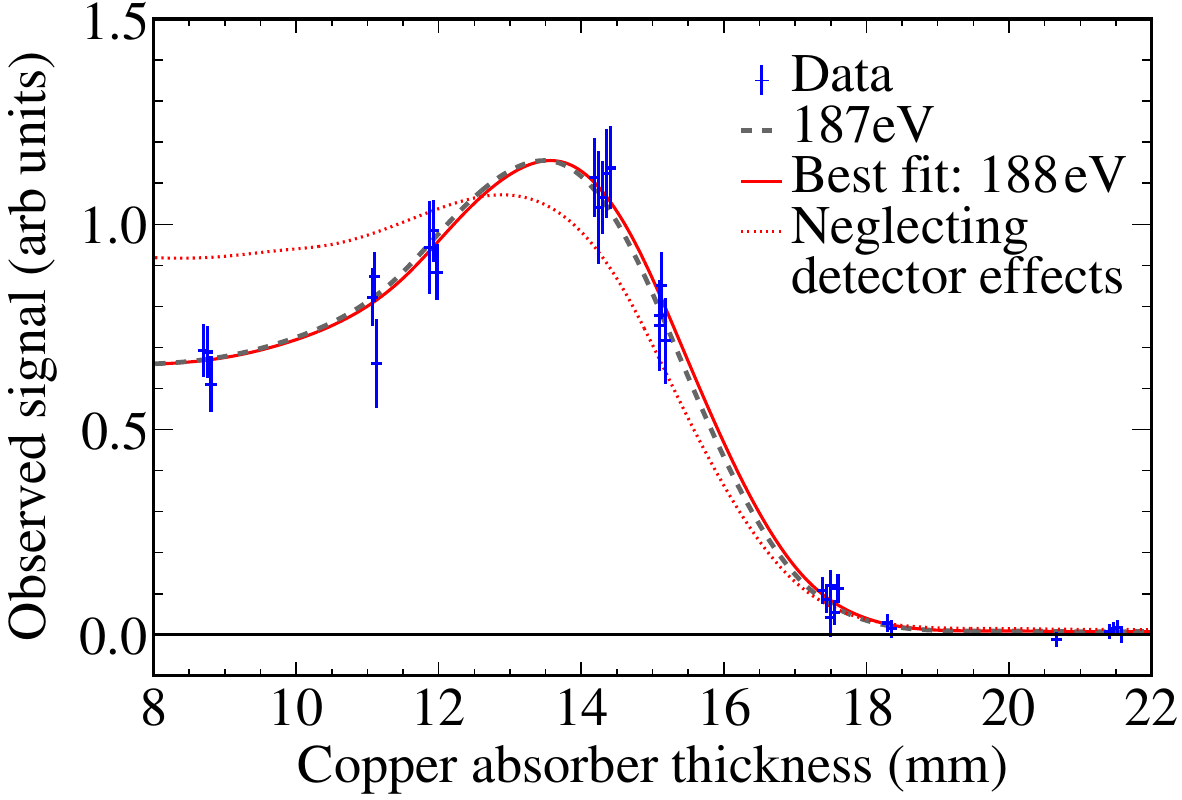}\hfill%
    \includegraphics[width=0.49\columnwidth]{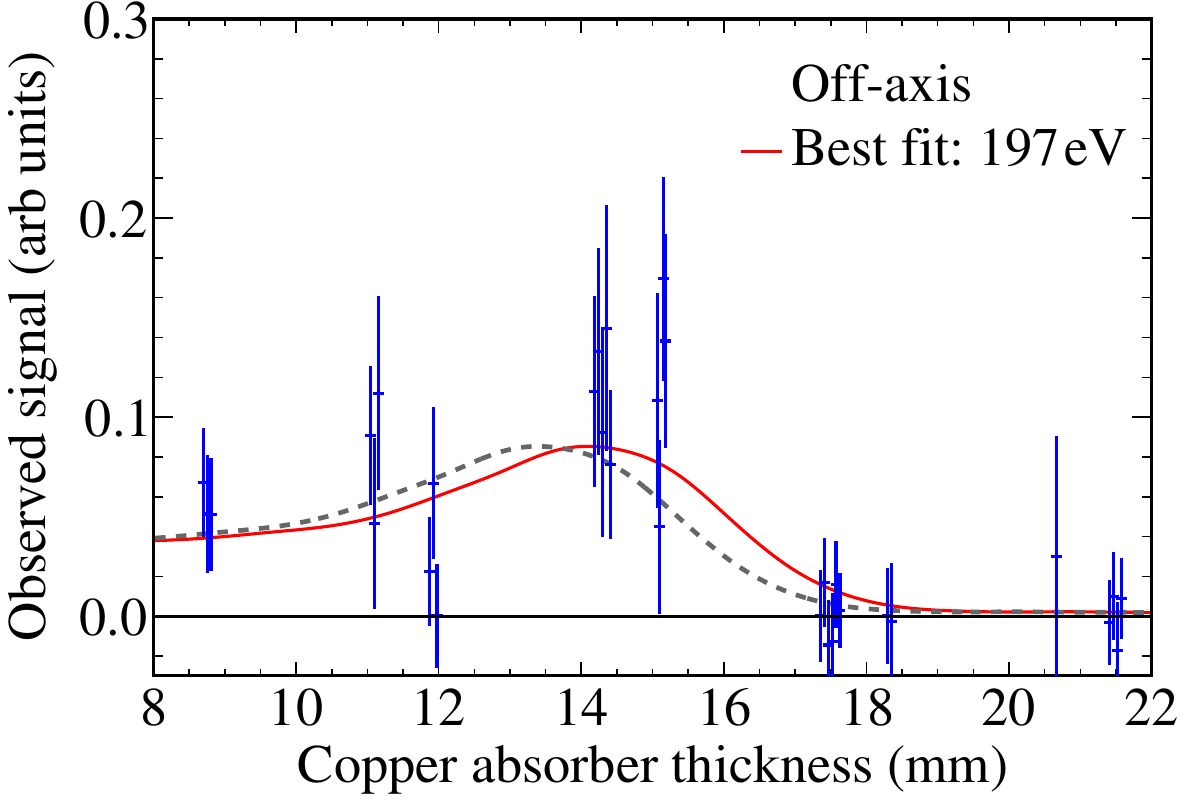}

    \caption{Left: Experimental data for on-axis scintillator (crosses) with a comparison to \geant simulations using
             $I=187$\,eV (the ICRU-90 value for gaseous argon) and 188\,eV (the best fit value). The solid red curve shows
             the best fit including systematics effects of the scintillator and camera, while the dotted curve shows
             the expectation for 188\,eV if detector response were directly proportional to energy deposition. Each cluster of data points
             is for a single copper absorber thickness and have been given slight horizontal separations
             to display the uncertainties.  Right: the same for the off-axis scintillator, showing only 187\,eV and the best fit.}

    \label{fig:realdatafit}
  \end{figure}
}

\newcommand{\figureevidence}
{
  \begin{figure}
    \centering

    \includegraphics[width=0.4816\columnwidth]{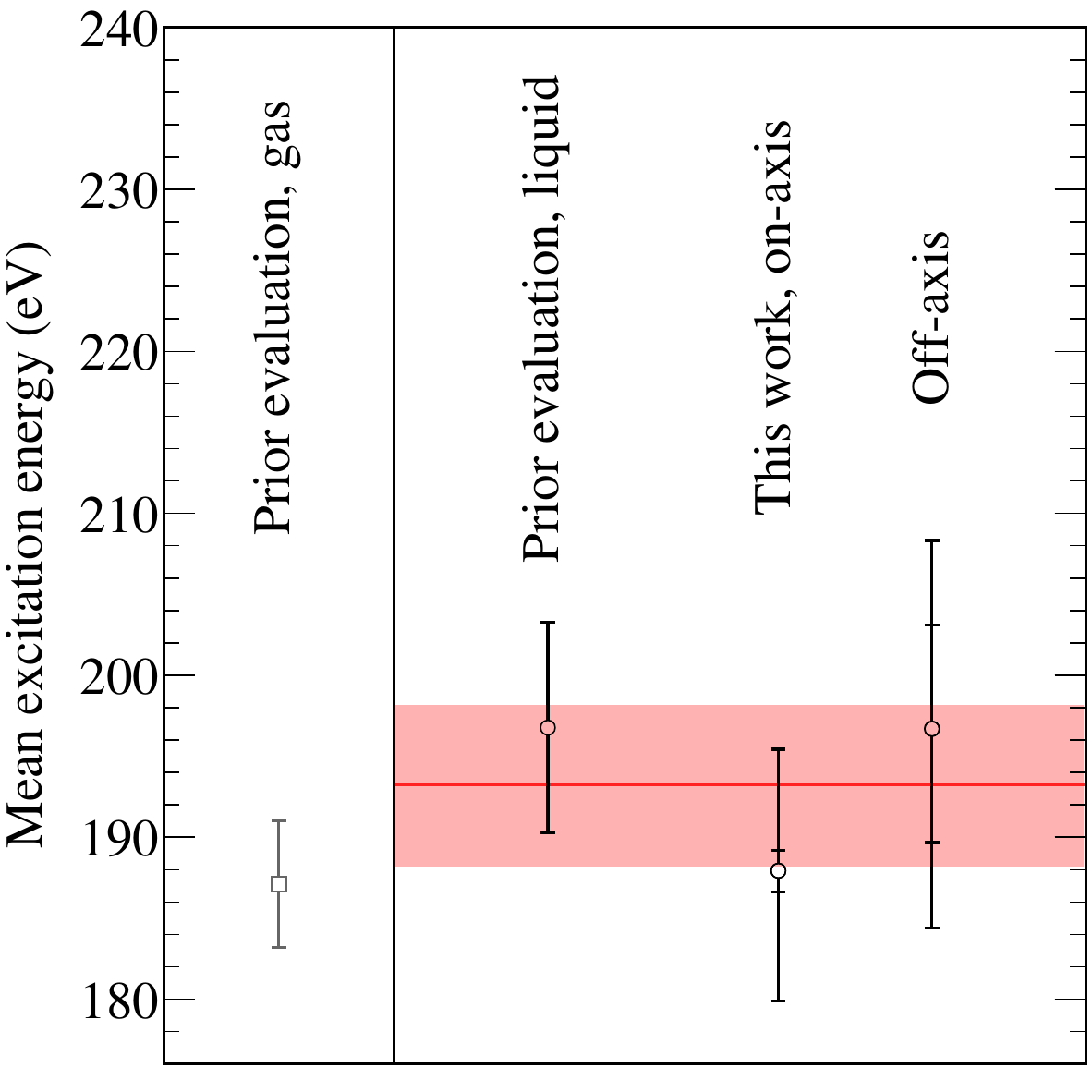}

    \caption{Combination of prior information with these results. The shaded
             band shows the result combining all past and present information.
             The present result for the on-axis and off-axis scintillators are shown
             on the right, with statistical and total errors.  }

    \label{fig:evidence}
  \end{figure}
}

\title{Evaluation of the mean excitation energy of liquid argon}

\author{M.~Strait}
\thanks{mstrait@fnal.gov}
\affiliation{PO Box 500, Batavia, Illinois 60150, USA}


\section{Introduction}

Neutrino detectors such as DUNE~\cite{dunetdr} depend on knowledge of the mean excitation
energy, also known as the \emph\ivalue, of liquid
argon to estimate neutrino energy.  The \ivalue controls the stopping power of
a substance for charged particles in the energy range where the Bethe formula is 
valid, $0.1 \lesssim \beta\gamma \lesssim 1000$~\cite{pdg}.  In a liquid argon neutrino detector, the
primary method to determine muon energy is by range, and the primary method of
calibrating the calorimetric response is to use the minimum ionizing portion of
muon tracks.  Each of these relies on correct simulation that depends on liquid
argon's \ivalue. Typically there are no calibration points in the GeV
range that can be used to mitigate an incorrect \ivalue.  While near detectors cancel out
neutrino flux and neutrino cross section uncertainties in long-baseline
analyses, they do not cancel out energy reconstruction uncertainties.

ICRU Report 90~\cite{icru90} gave the I-value of gaseous argon as $(187\pm 3)$\,eV. 
The present author has re-evaluated the world data on gaseous argon and recommends
the same central value with a larger uncertainty: \garanswer~\cite{self}.  No previous
recommendation had been given for liquid argon, with mainstream
simulation packages such as \geant assuming the same \ivalue for all phases, and using
the older value of 188\,eV from ICRU Report 37~\cite{icru37}.  In Ref.~\cite{self},
the author recommended \laranswer, based on very sparse solid argon data and an extrapolation
from other substances whose stopping power has been measured in both the gaseous 
and condensed phases.  The stopping power of liquid argon has not previously
been measured.

We report on the first measurement of \ivalue of liquid argon.
Sections~\ref{sec:experimental} and \ref{sec:data} describe the experimental setup and data set.
Section~\ref{sec:fit}
explains the fitting procedure, while \sect{systematics} presents
an analysis of the systematic uncertainties.  Section~\ref{sec:conclusions}
gives our recommendation for the \ivalue and concludes.

\section{\label{sec:experimental} Experimental setup}

\figurescheme

A vessel containing liquid argon was exposed to the Fermilab Linac beam at the
Irradiation Test Area (also known as the MuCool Test Area).  See schematic and
photograph in \fig{scheme}.  The beam is at a fixed energy, and is pulsed.  For
this measurement, pulses came once a minute with a length of 7\,$\mu$s, each
pulse consisting of $7\times 10^{11}$ protons.

The primary beam from the Fermilab Linac is $\mathrm H^-$ with a central energy
of $(404.0\pm0.8)$\,MeV, which, after stripping by a 0.05\,mm titanium beam
window, gives $(403.6\pm0.8)$\,MeV protons.  The beam kinetic energy has an
approximately Gaussian distribution with an RMS of
$(0.50\pm0.05)$\,MeV~\cite{Bhat:2019raz}.  The stripped electrons, with kinetic
energy of 220\,keV, are rapidly stopped at the upstream end of the experimental
setup. Simulation shows that their secondaries contribute only on order of
$10^{-7}$ as much energy in the downstream detectors as the protons.  They are
neglected from here on.  The beam profile was monitored by a wire chamber
upstream of the vessel.  It is close to Gaussian with an RMS of 1.2\,cm both
horizontally and vertically.  The beam pipe is 10\,cm in diameter.

The thickness of the liquid argon and its vessel in the beam direction was
chosen to be somewhat less than enough to stop the protons.  The argon vessel
was a rectangular open-topped aluminum 6061 tub with interior length, in the
beam direction, at 87\,K, of 654.1\,mm (656.6\,mm at room temperature).  The liquid argon was obtained from
Airgas and was industrial grade, 99.997\% pure.  The interior width was 220\,mm
and the height 160\,mm.  The walls of the vessel were 17.76\,mm thick.  The
vessel was surrounded by Foamular-250 polystyrene insulation boards, with
51\,mm at the front and back in the beam direction, 178\,mm on the right and
left sides, 203\,mm below and 229\,mm above.  The insulation was
sufficient to keep an adequate volume of liquid argon for many hours at a time
without active cooling.  The insulation was wrapped in aluminum foil for fire
protection, but not the section directly in the path of the beam.

Immediately upstream of the argon vessel were two fixed copper strips with total
thickness 6.33\,mm.  Their purpose is to reduce the beam energy to better match the
size of the liquid argon vessel.  A copper assembly 0.5\,m further upstream contained 20
thicknesses of copper, evenly spaced from zero to 15.17\,mm, each of which could be
remotely moved into the beam to scan over beam energies.  In this way, the mean proton
energy entering the vessel's front face could be adjusted from 362\,MeV (the greatest
copper thickness) to
391\,MeV (the least copper thickness).  The Bragg peak and subsequent rapid decrease in
transmittance is observed as a function of increasing copper thickness.

The transmittance of the materials in the beam was measured by two
ceramic scintillators 0.5\,m downstream of the vessel, one directly
on the beam axis and the other 14\,cm off-axis.  Made by Advatech, the
scintillators are 50\,mm by 50\,mm by 1\,mm.  They are 99.5\% aluminum(III) oxide
doped with chromium(III) oxide, known as \emph{chromox}, with density 3.85\,g/cm$^3$~\cite{MuonAccelerationRD:2013mbu}.  Both scintillators
were observed by a single commodity video camera, a Logitech C920, positioned 1.2\,m
off the beam axis.  The camera was set to manual exposure mode with a fixed
30 frames-per-second readout.  The focus was fixed and set manually.  Likewise,
automatic white balancing and backlight compensation were disabled.

The scintillators were angled 30$^\circ$ towards 
the camera with respect to the beam axis.  Immediately downstream of
the vessel is a stack of concrete shielding blocks with a ``cave'' 1\,m
wide by 1\,m tall along the beam path that leads to the beam dump.  The
scintillators were positioned inside this cave, which, compared to an open
configuration, slightly increases the flux of secondary particles they
observe along with the primary protons.

\section{\label{sec:data} Data collection}

Data was obtained for 9 thicknesses of copper across a total of 31 pulses,
with between 1 and 6 pulses per thickness. See Table~\ref{tab:order}. Unforeseen circumstances prevented
further data taking. A more even distribution of data
across all 20 thicknesses of copper would have been desirable, but this data is
sufficient to make a measurement.  

\tableorder

Camera images were analyzed to determine the relative amount of ionization in the
scintillator from each pulse.  As shown in \fig{images}, images contained both
scintillation light and background from secondary beam particles directly striking the
camera.  The former registers almost exclusively in the camera's red pixels,
while the latter appears equally in all three color channels, since
radiation directly ionizes the CMOS sensors, bypassing the color filters.  By
subtracting the signal in blue and green pixels from that in the red pixels,
the scintillation light is isolated.  Empirically, subtracting the green
channel scaled by 0.44 and the blue channel scaled by 0.59 yielded the best
results, as judged by regions of the images away from the scintillator.
The resulting pixel values (including those that are negative after subtraction) are then summed for each
scintillator panel.

The camera gain was set to avoid saturation in any of the pixels for the
on-axis scintillator.  However, this meant that the signal observed in the
off-axis scintillator was very weak.  The off-axis data is nevertheless
sufficient to cross-check the main result from the on-axis scintillator.

As the scintillator emission time is comparable to the camera frame rate, 30
frames per second, the signal from the frame containing the pulse and the
following frame were summed.  The following frame is free of beam background,
but only contains a small minority of the light, so cannot be used alone to
avoid the background.

Differences in the measured signal from pulse to pulse can be caused by
variations in the beam intensity, the number of particles striking the camera
in the pixels observing the scintillator, and effects related to the camera
readout.  These effects are not otherwise accounted for, and collectively will
be referred to as statistical uncertainty.  The statistical uncertainty has
been estimated via the spread of measurements using the same copper absorber
thicknesses.  It is strongly subdominant to the systematic errors that will be
discussed in the following sections.

\figureimages

\section{\label{sec:fit} Fit procedure}

The experimental setup was simulated using \geant v11.2.1~\cite{geant2016}. 
See \fig{geant}.  The physics list
\texttt{FTFP\_BERT\_EMZ} was used, where \texttt{EMZ} is also known as EM
\texttt{Opt4}, and provides the best EM physics available in \geant.\footnote{https://geant4-userdoc.web.cern.ch/UsersGuides/PhysicsListGuide/html/electromagnetic/}
For all materials, the \texttt{DensityEffectCalculatorFlag} was set to obtain
the best available treatment of the Fermi density effect. At the relevant
energies, this has no effect for argon but a small effect for the aluminum
vessel and copper strips, since they have more loosely bound electrons.

\figuregeant

The experiment was simulated with the \ivalue of liquid argon set to 181\,eV,
189\,eV, 197\,eV, 205\,eV, and 213\,eV and with each of the 20 copper absorber
thicknesses, with $10^7$ to $10^8$ protons simulated in each case, with higher
statistics used for points providing more information to the fit, such that the
uncertainty from Monte Carlo statistics is negligible.  The normalization of
the signal is a free parameter in the fit; i.e. we do not attempt any a~priori
calibration of the scintillator yield or camera's efficiency.  As a very good
approximation, the transmission as a function of copper absorber thickness
differs from one \ivalue to another only by a horizontal translation and a
normalization (see \fig{shape}).  Therefore, the shape of this curve is used as
a template in the fit which extracts the experimental \ivalue from the data. In
the fit, the parameter of interest is the left-right shift (absorber thickness).
Additional parameters are the up-down shift
(background normalization), the overall normalization, and three free parameters for
several systematic effects, discussed below.  Because the off-axis data has large
statistical uncertainties, the background normalization is fixed 
to the Monte Carlo prediction, and three systematics parameters are fixed to the best
fit values found in the on-axis fit.  The results are shown in \fig{realdatafit}.

\figureshape

\figurerealdatafit

It was found that the on-axis data prefers a sharper Bragg peak than the
prediction under the naive assumption that the detector signal is simply
proportional to energy deposition.  This is particularly pronounced with the
data taken at 8.8\,mm, although the data at other positions also tends to
prefer a sharper peak.  We investigated whether the 8.8\,mm data could be
flawed, but did not find any evidence of this.  For instance, it was not taken
first or last (see Table~\ref{tab:order}), and the data from pulses immediately before and after it
(11.9\,mm and 15.1\,mm, respectively) are not similarly lower than expected.
Beam monitoring did not show the 8.8\,mm pulses to be abnormal, and we could
independently confirm this from the magnitude of the beam background observed
in the obtained images.  

We investigated several detector effects and determined that the observed shape
could be explained via a combination of quenching in the scintillator and
different emission spectrum of the scintillator as a function of particle
dE/dx, convoluted with the camera's spectral response.  These effects, and systematic uncertainties associated with them, will
be discussed at length in sections~\ref{sec:scintillator} and \ref{sec:beam}.

The statistically independent results of $(188 \pm 1\mathrm{(stat)})$\,eV and $(197^{+6}_{-7}
\mathrm{(stat)})$\,eV from the on-axis and off-axis scintillators,
respectively, are in good agreement.  The statistical uncertainty is defined as
the shift necessary to produce a $\Delta\chi^2$ of 1.0 when only the mean ionization
energy and the overall normalization are allowed to vary.

\section{Systematic uncertainties} \label{sec:systematics}
Many sources of systematic error were investigated.  All of the uncertainties discussed
in this section are summarized in Table~\ref{tab:syst}.

\tablesyst

\subsection{Beam}\label{sec:beam}

The design $\mathrm{H^-}$ energy of the Linac is
401.5\,MeV~\cite{Schmidt:1993fz}.  We estimate the present ${\mathrm H^-}$ energy to be
$(404.0\pm0.8)$\,MeV, i.e. a proton energy of $(403.6\pm0.8)$\,MeV.  This
estimate derives from three sources of information.  First, the beam energy was
determined by using the known circumference and RF frequency of the Booster,
which receives protons from the Linac, and cross-checked by further
characteristics of the accelerator chain.  This led to an estimate for the
proton energy of $(402.2\pm0.2)$\,MeV.  A second estimate was made of
$(403.2\pm0.1)$\,MeV, based on the time of flight of Linac beam injected into
the Booster.  While these estimates are in conflict, neither could be
identified as superior to the other.  They may indicate drift in the beam
energy, which is known to occur at the level of 0.5\,MeV to 1.0\,MeV. The relative drift was not
monitored between the two studies just described, nor from the studies to
the time of data taking for the present measurement.

Additionally, a later transmission measurement for copper using the same technique as we 
report for liquid argon suggested a proton energy of 404.5\,MeV, assuming the 
adopted I-value for copper.  Again, this difference could indicate either real drift
in the beam energy or that one or more of these techniques for estimating the energy is
incorrect.  As the first two energy estimates are correlated through the assumed circumference 
of the Booster, we take their mean and then average this with the estimate from the
copper transmission measurement, arriving at $(403.6\pm0.8)$\,MeV.
This uncertainty leads to an
uncertainty on the \ivalue of 6\,eV.

In contrast, the measured I-value is quite insensitive to the
details of the energy spread, with a shift of less than 0.1\,eV resulting from
the $(0.5\pm0.05)$\,MeV uncertainty on the beam energy width.  It is notable
that nearly all of the spread of the Bragg peak is
caused by range straggling in the experimental materials and not the initial
energy profile of the beam.
The details of the energy profile are likewise not important; for instance if
the profile, instead of a single Gaussian with an RMS of 0.5\,MeV, were a sum
of two equally weighted Gaussians, one with RMS 0.5\,MeV and one with RMS
2.0\,MeV, the result for the I-value shifts by only 0.12\,eV.
The nominal beam spot size is 1.2\,cm RMS, verified by beam monitors during the
run.  Even large changes to this size have a negligible effect on the inferred
\ivalue.

\subsection{Scintillator and camera}\label{sec:scintillator}

According to the manufacturer and
CERN-PS-90-42-AR~\cite{johnson1990development}, chromox scintillator has an
extinction coefficient of $(8\pm 1)$\,cm$^{-1}$ for its peak light production
frequency of 693\,nm.  A more recent study by
Cook, Tresca and Lefferts 2014~\cite{Cook_2014}
measured the attenuation
length using transmission of 633\,nm light --- substantially below the
peak emission --- to be $(0.05\pm0.01)$\,cm, i.e. an
extinction coefficient of $(20\pm 4)$\,cm$^{-1}$, and also
found that including a Birks' quenching factor gives a better fit to proton
irradiation data.  Their best fit was for a Birks' constant
of $2.33\times 10^{-2}\,\mathrm{g\,MeV^{-1}\,cm^{-2}}$ and attenuation length of
$0.058$\,cm or $0.057$\,cm (both values are quoted).  This quenching
constant is large compared to typical values for inorganic scintillators,
although not outside what has been measured before~\cite{Adriani}.

The effect of scintillator quenching is to flatten the Bragg peak, since
quenching suppresses light production from low energy protons.  This is substantially
at odds with our data, which prefers a sharper Bragg peak than the simulation predicts
even without quenching.  A resimulation of Cook's setup was performed to investigate
this discrepancy.  Using the data from their Figure~2, we find that
a much smaller Birks' constant along with a shorter attenuation length fits
the data slightly better (see \fig{cook}).  Since they viewed their scintillator from the back,
both parameters result in loss of light for protons with a range shorter than
the scintillator thickness, and so the two effects cannot be readily
separated.  On the other hand, both their direct attenuation measurement
and the manufacturer's statement disfavor a shorter length.

\figurecook

We choose to take
$(2.33\pm0.50)\times 10^{-2}\,\mathrm{g\,MeV^{-1}\,cm^{-2}}$ as the nominal value and uncertainty of Birks'
constant.
We estimate this uncertainty from the amount of freedom in the quenching allowed by Cook's attenuation
measurement.  The uncertainty in the \ivalue incurred by the quenching turns out to be negligible
because its effect is relatively well-constrained and, in any case, it is degenerate with the effect we will
discuss next.

To explain the discrepancy between our data and the expected shape of the Bragg peak, we
suggest that the scintillation spectrum of chromox shifts with the dE/dx of
protons.  Such a shift has been observed when comparing emission from x-ray and alpha
irradiation in zinc oxide and zinc sulfide scintillators~\cite{ZnO}, in which the highly
ionizing alphas result in a spectrum shifted roughly 10\,nm down.
Since we observed the chromox emission through both an infrared filter (known as a ``hot mirror'' by
photographers) and a red filter, both part of the camera,
we are sensitive to the integral of the product of the emission spectrum and transmission
efficiency. A similar 10\,nm shift would cause a 50\% change in the signal (see \fig{shiftchromox}) because of
the sharp edge of the infrared filter, while a 20\,nm shift would cause a 200\% change.

\figurechromoxspectrum

A second effect that may skew the signal as a function of dE/dx is a varying scintillator
decay time.  It has been observed that x-rays produce long afterglows in chromox~\cite{chromox-xray}
and also that the decay time when irradiated with 50\,keV He$^+$ is much longer (1.6\,s to fall 90\%)
than when irradiated with 1.06\,MeV protons (72\,ms) or 1.65\,MeV He$^+$ (60\,ms)~\cite{chromox-hhe}, all of which
are much larger than the manufacturer's figure of 3.4\,ms, which perhaps refers to the
decay time for high energy proton irradiation.  

Since we lack specific data for either the spectral shift or the scintillator decay
time as a function of dE/dx, we make a generic assumption
that there is one efficiency for low-dE/dx particles and another for high-dE/dx
particles, with some transition region between these.  We allow the fit to
determine the cutoff dE/dx, the width of the transition, and the ratio of
efficiencies.  We limit the ratio of efficiencies, low-dE/dx:high-dE/dx, to be between 1:1 and 1:3, corresponding
to a spectral shift in the chromox of between zero and 20\,nm.  The data has relatively little power to simultaneously constrain
all three parameters, leading to a large systematic uncertainty.  The fit prefers an abrupt
transition at 60\,MeV/cm with an efficiency ratio of 1:2.3, but given the degeneracies
between parameters, these specifics should not be taken too literally.

The effect of higher efficiency for higher dE/dx is to increase the signal on the falling
edge of the Bragg peak while lowering it on the leading edge.  This is illustrated by the
difference between dotted and solid lines in \fig{realdatafit}, left.  The change allows
both the data in the peak and that at 8.8\,mm to be well-fit.  Unfortunately, it is
somewhat correlated to an overall right-left shift and so incurs a large systematic
uncertainty of $^{+3.2}_{-4}$\,eV.

Finally, we take a systematic uncertainty for the attenuation length itself.
The attenuation lengths from the manufacturer of 0.125\,cm for 693\,nm and
from Cook of 0.05\,cm for 633\,nm may both be correct, since similar
materials are known to have rapidly changing attenuation lengths in this
wavelength region~\cite{rubyatten}, with shorter lengths at shorter wavelengths.
As can be seen from \fig{shiftchromox}, the visible signal for us is
split roughly evenly between the sharp peak at 693\,nm and the broad
peak at 620--680\,nm.  The relative contributions are not well known
because of the possible spectral shift effect discussed above,
nor do we know how the attenuation length evolves between the two
measured wavelengths.

Changes in the scintillator attenuation length are almost exactly degenerate
with an overall left-right shift of the predicted light curve, as it is
equivalent to adding or subtracting material in the protons' path.  For
example, for an arbitrarily short attenuation length, any proton that reaches
the front face produces a visible signal; protons that continue farther
produce no additional signal.  At the other extreme, for an arbitrarily long
attenuation length, light production from all depths contributes, meaning
protons with enough energy to reach the back face produce the greatest signal.
This situation is nearly equivalent to having a half-thickness of scintillator
of extra material in the protons' path.  (The attenuation length also changes
the overall normalization of the signal, but this is a free parameter in our
fit, and so after fitting, only the left-right shift remains.)
Wavelength-dependent attenuation may introduce an additional term into
the efficiency shown in \fig{shiftchromox}, but no such detailed
information is available.  Since the fit described above is done using
a generic model without direct use of wavelength-dependent details, this additional
term has no effect on the analysis.  We conservatively consider the extreme cases of all of the light having an attenuation length
of 0.05\,cm and all of it having an attenuation length of 0.125\,cm. The
difference in inferred \ivalue between these extremes is 0.7\,eV.  We take the
midpoint as the nominal result, assume a uniform PDF over the range and claim
an uncertainty of 0.2\,eV.

\subsection{Materials accounting}\label{sec:matcount}

\subsubsection{Aluminum vessel}

The front and back walls of the aluminum tub are $(17.83\pm0.03)$\,mm thick at
room temperature.  At the boiling point of argon, 6061 aluminum linearly
contracts by a factor of $0.99621\pm0.00015$~\cite{Marquardt2002} to
$(17.76\pm0.03)$\,mm.  Given the thermal conductivities of the
insulation and aluminum, the aluminum's temperature is
uniformly within 0.1\,K of the argon temperature.  The uncertainty associated
with the amount of contraction is negligible compared to the uncertainty of the
thickness at room temperature.  The uncertainty in the total thickness of
aluminum of 0.06\,mm has 35\% the stopping power of the same thickness of
copper.  From the Monte Carlo simulation described above, each additional
millimeter of copper degrader needed to stop the beam corresponds to an
additional 15\,eV in liquid argon's \ivalue.  The uncertainty in aluminum
thickness therefore gives an uncertainty in the \ivalue of 0.3\,eV.

Given the specification of 6061 aluminum,
there is some freedom in its elemental composition. The first-order effect for
the stopping power is the material's mean $Z/A$.  The allowed range is 0.12\%,
giving an uncertainty in liquid argon's \ivalue of 0.06\,eV.  The density of
our particular aluminum was not measured.  We take 6061 aluminum density to be
$(2.69\pm0.01)$\,g/cm$^3$~\cite{al6061} at room temperature --- $(2.72\pm0.01)$\,g/cm$^3$ 
at 87\,K --- and further assume the density uncertainty is uncorrelated to the
uncertainty in its elemental composition, leading to a total uncertainty of 0.6\,eV.
The interior length of the aluminum vessel has a tolerance of 0.25\,mm.  The
resulting uncertainty in the thickness of liquid argon leads to an \ivalue
uncertainty of 0.3\,eV.  The aluminum's density dominates the uncertainty from
the aluminum vessel, which is 0.8\,eV in total.

\subsubsection{Copper strips}

The density of the copper strips is taken to be $(8.94\pm0.02)$\,g/cm$^3$,
using a combination of public data on density of copper and our own
measurements of copper blocks from the same supplier.  The copper's temperature
was the temperature of the air in the hall, $13^\circ$C.  The difference in its
density at this temperature and at the nominal $20^\circ$C is negligible.  Each
strip's thickness is known to $\pm0.008$\,mm, and we conservatively assume full
correlation of this uncertainty from strip to strip.  The resulting
uncertainties on the I-value of liquid argon from density and thickness are
0.5\,eV and 0.3\,eV, respectively.

\subsubsection{Insulation}

The insulation is Foamular-250 polystyrene, with a manufacturer-quoted minimum density of
0.0248\,g/cm$^3$.  Taking the insulation to have $(0.027\pm0.002)$\,g/cm$^3$
leads to an I-value uncertainty of 0.4\,eV. 

The inner surface of the insulation is at the temperature of the aluminum
vessel, approximately the boiling point of liquid argon, while the outer
surface is at the ambient 13$^\circ$C.  The thermal contraction of the
insulation was not analyzed in detail, since its change in density between the
ambient temperature and liquid argon temperature, 4\%, is less than the uncertainty
of its room temperature density.

\subsubsection{Density of liquid argon}

Because the argon was not actively cooled, it was at its boiling point, which
is significantly affected by the atmospheric pressure.  During the experimental
run, the sea-level pressure recorded by the weather station at the O'Hare
airport 33\,km to the east-northeast was 1004\,hPa and by the Dekalb/Cortland airport 38\,km to the
west-northwest was 1005\,hPa.  Other area weather stations agreed within 1\,hPa.  Taking
the mean of these, and converting from the reported sea-level pressures to the
pressure at the experiment's elevation, 225\,m, gives
978\,hPa.  The uncertainty on the readings from each weather station is
conservatively taken to be 0.7\,hPa, corresponding to the required $\pm
0.02$\,inHg accuracy required by the United States Federal Meteorological
Handbook No.~1, 2019~\cite{handbook}.  The accuracy of converting the sea level
pressure to the local pressure at Fermilab's altitude is better than
1\,hPa~\cite{SeaLevelPressure}.  The vessel was unpressurized, with insulation
placed loosely on the top, such that the pressure difference between the
interior and exterior was much less than 1\,hPa.  Overall, we take the pressure
to be $(978\pm1)$\,hPa.

From Ref.~\cite{argondensity}, given the pressure, the argon's boiling point was
$(86.97\pm0.02)$\,K, where the uncertainty of the pressure dominates over the
uncertainty of the correspondence between the pressure and the boiling point,
given a 0.04\% uncertainty in the vapor pressure as a function of temperature.
This is significantly lower than argon's standard boiling point of 87.303\,K.

Correspondingly, again using Ref.~\cite{argondensity}, its density is
$(1.3975\pm0.0004)$\,g/cm$^3$, above the density at standard pressure of
1.3953\,g/cm$^3$.  The uncertainty has equal contributions from the uncertainty
in the boiling point and the uncertainty of $\pm 0.02\%$ from the
correspondence between temperature and density.  The density uncertainty
results in an I-value uncertainty of 0.4\,eV.

\subsubsection{Other material adjustments}

The vessel was filled with liquid argon with a total depth of 16\,cm.  Data was
taken between two and four hours after the fill.  Previous tests without beam showed
that the loss rate of liquid argon is 0.3\,cm per hour.  Each beam
pulse adds additional heat, but this effect increases the evaporation rate
by only 0.1\%.  Therefore, we estimate that the argon level
varied from 15.4\,cm to 14.8\,cm during the experimental run.  Monte Carlo
simulations used for the fit assumed a depth of 12\,cm, but we find no
significant differences in simulation results for depths between 11\,cm and 16\,cm.   Even at 11\,cm,
there is continuous argon for all straight-line paths from the beam pipe to
the scintillator, but small modifications in transmission begin to occur because
protons which are scattered upwards into the air have no significant chance of being
scattered back down towards the scintillator.

Although the liquid argon is at its boiling point, the insulation was chosen to
ensure that evaporation would occur only at the surface of the liquid, and not
through nucleate boiling, meaning that the path the protons traverse contains
only argon in the liquid phase.  Tests of the vessel with liquid nitrogen
showed only very limited nucleate boiling, even with the lid removed and the top
of the nitrogen exposed to room temperature air.  Liquid argon has a higher
density, higher boiling point and larger heat of vaporization, all of which
reduce its boiling rate relative to nitrogen.  Given these observations,
we estimate that the protons' mean path length through gaseous argon is no
greater than 0.01\,mm.  This being much smaller than the 0.25\,mm tolerance on the
vessel's interior length, it has no effect on the measurement.

We also consider the possibility that a thin layer of frost formed between the
insulation and the aluminum vessel.  During tests when the vessel was
constructed, there was no indication of this problem occurring, but it was not
possible to inspect the vessel after the beam run.  Given the degree of
flushness between the aluminum vessel and the insulation, we consider 0.25\,mm
to be the largest credible thickness of frost.  We model the mean case in the
simulation, i.e. 0.125\,mm of ice on the front and back of the vessel.  An
uncertainty of 0.2\,eV is incurred from the range of possible frost
thicknesses.

\subsection{Multiple Coulomb scattering}\label{sec:mcs}

Multiple Coulomb scattering affects the projected range of protons through the
liquid argon.  Our analysis uses \geant's electromagnetic \texttt{EMZ} option,
also known as \texttt{Opt4}.
This option uses the best available multiple scattering model in
\geant: ``multiple Coulomb scattering is performed by the \texttt{WentzelVI}
model and Coulomb scattering by the \texttt{eCoulombScattering} model.''
The \texttt{WentzelVI} model implements the
algorithm from Fern\'andez-Varea 1993~\cite{fernandezvarea}, using the
potential form introduced by Wentzel 1926~\cite{wentzel}.  This remains an
approximation, for instance there are no detailed models of atoms' potentials,
only the simple Wentzel form \[V = \frac{e E}{r}e^{-r/R},\] where the atomic
radius is taken as a continuous function of $Z$: \[ R = 0.885
Z^{-1/3}a_0\left[1.13 + 3.76(\alpha Z/\beta)^2\right]^{-1/2}.\]

We compared \texttt{EMZ} to \geant's single scattering model, in which
every Coulomb scatter is simulated independently instead of using an
approximate distribution to simultaneously simulate the effect of many
scatters.  This is enabled using the \texttt{FTFP\_BERT\_EMZ} physics list,
followed by \texttt{ReplacePhysics(new G4EmStandardPhysicsSS)}.  The single
scattering model is extremely CPU intensive to run, being three orders of
magnitude slower than multiple scattering.  Only one degrader thickness was
tested, 15.9\,mm, which is the most sensitive to differences in proton range.
The results were compatible with EMZ, with single scattering giving a signal
$(1.6 \pm 1.2)$\% larger than EMZ.  This is equivalent to a decrease of the
\ivalue by $(0.3\pm0.2)$\,eV, assuming the shape of the complete transmission
curve is the same for the two models.

While the single scattering model is naively the most accurate, and has been shown
to be so for thin targets, it is difficult
to show experimentally for thick targets~\cite{Ivanchenko:2010zz,SOTI201311,Makarova_2017}.
Given that, and the prohibitively high CPU time needed for a full simulation using it,
we retain EMZ as our model for this analysis, as it is the most accurate model that
is practical to use.
Were the single scattering model the ground truth, it would be reasonable to
assign a 0.5\,eV systematic error associated with using EMZ instead.  However,
acknowledging that even the single scattering model still makes use of
several approximations, we choose to double this to 1.0\,eV.

We also checked the effect of using the default, less accurate, 
\texttt{Opt0} electromagnetic option from the \texttt{FTFP\_BERT} physics list,
which uses the so-called \texttt{Urban} model for multiple
Coulomb scattering.  The \texttt{Urban} model uses the Highland approximation for multiple
scattering, \[\theta_\mathrm{rms,\ plane} = \frac{13.6\,\mathrm{MeV}}{p\beta}
z\sqrt{x/X_0}[1+0.038\log(x/X_0)],\] which is rapid to compute at
the expense of accuracy.   It predicts a longer projected range than \texttt{EMZ}
for the same \ivalue, such that a fit to the data using \texttt{Opt0} predictions
results in an inferred \ivalue 5.5\,eV lower than that inferred using \texttt{EMZ}.
We do not use this rather different result except to note that
experiments which rely on absolute particle ranges should certainly use \texttt{EMZ}
and not the default \texttt{Opt0}.

We studied whether a larger or smaller scintillator panel would have resulted
in less sensitivity to the details of multiple scattering models.  The differences
are largest along the beam axis, but not dramatically.  Using a scintillator
of arbitrarily large transverse extent reduces the differences between Opt0 and
Opt4/\texttt{EMZ} by only 5\%.  For a scintillator panel of width 1\,cm the difference
is 30\% larger than for the 5\,cm panel.  This study suggests that the
primary difference in the models is in the projected range rather than the
angular distribution.

\subsection{Monte Carlo code consistency}\label{sec:mcconsistency}

We reran the simulations using \textsc{mars} version 2024-04-25~\cite{mars}, CERN FLUKA
version 4-4.0~\cite{fluka}, and PHITS version 3.34~\cite{phits}. See \fig{mars}. \textsc{Mars} was run
in \textsc{mcnp} mode for neutron transport below 14.5\,MeV, and otherwise in
its default configuration. The hadronic cross section was reduced by 5\% to better match \geant. A ``Z-sandwich'' geometry was used, modeling
the experimental design with a series of simple slabs; this is a good
approximation given the large transverse extents of the experimental
components, and allows for much better code performance.  The dose on the
scintillator scored in a number of radial layers and a weighted sum was used to
determine the dose for the true geometry.  FLUKA was run in \textsc{precision}
mode.  PHITS was run with Landau-Vavilov energy
straggling (\texttt{nedisp = 1}), the NMTC model~\cite{NMTC} for multiple Coulomb
scattering (\texttt{nspred = 1}), the ATIMA model for
energy loss (\texttt{ndedx = 1}), and using \texttt{e-mode = 1} for
low-energy neutrons.

\figuremars

The other Monte Carlo codes predict slightly shorter or longer proton ranges
than \geant given the same \ivalue, meaning that analyzing the data using them
results in, respectively, higher or lower \ivalues.  FLUKA and \textsc{mars} each give
a result 1.6\,eV higher than \geant, while the answer from PHITS was 1.0\,eV
lower.  As is visible from the plot, the largest difference between the codes is
on the left side, before the peak, although it is the falling edge on the right
side which gives the range.  The differences between codes to the left of the peak
are much smaller than the detector effects discussed in \sect{scintillator}.

Differences between Monte Carlo codes can result from such a wide variety of
causes that it is not easy to pin down the source.  In each case, we used the
most accurate options available so far as our expertise allowed.  It can be
seen that the difference between various codes is similar to the difference
obtained by using different options within \geant, e.g. the multiple scattering
model.  It is reasonable to suppose that similar choices between several
available approximations is the reason for the differences from one code to
another.  We conservatively choose to take the RMS of the results from the four
codes, 0.9\,eV, as an additional systematic uncertainty, although this may represent
double counting to some extent.  The central value obtained from \geant is
retained as our central value, however if the reader would prefer to use the
mean of the four codes, add 0.7\,eV to our stated experimental result.

\subsection{Other \ivalues}\label{sec:otheri}

Using values from ICRU-37, and interpreting them as being at 90\% C.L. as suggested
by footnote~10, the \ivalue of the 6061 aluminum alloy is $(167.0\pm 1.2)$\,eV.  The resulting
uncertainty in the stopping power of the vessel leads to an uncertainty in
liquid argon's \ivalue of 0.2\,eV.  Similarly, the uncertainty of copper's
I-value, $\pm 6\,$eV, leads to a 0.7\,eV uncertainty in this measurement.

\subsection{Other physics uncertainties}\label{sec:otherphysics}

\subsubsection{Hadronic cross sections}

While there is some uncertainty on the inelastic scattering cross sections of 
protons on argon nuclei, these interactions have almost no impact on the measurement, since
any proton that participates in such an interaction both tends to lose a great deal of energy,
meaning that it no longer can arrive at the far end of the vessel, and also tends to get scattered at a large angle
so that it would not reach the detector regardless. The effect of increasing
the hadronic interaction cross section is therefore almost entirely to decrease the normalization
of the signal, as a function of beam energy, without changing its shape significantly.  An adjustment
of 20\% from \geant's default cross section makes only a 0.1\,eV impact on the inferred argon I-value.

\subsubsection{Shell corrections}

Shell corrections to the Bethe stopping power formula become important at low proton energy. 
These are difficult to calculate precisely, but are a minor effect for 400\,MeV protons.
For argon, a low-$Z$ nucleus, the uncertainty on the magnitude of the correction is under
10\%~\cite{icru37}.  A change in the correction at this level has only a 0.3\,eV impact on the
\ivalue.

\subsubsection{Delayed energy}

For each pulse, we integrate the scintillation light collected for two camera
frames, 66.7\,ms, which means that there is some sensitivity to delayed
processes, including neutron capture and decay of activated materials.
However, simulation shows that even when the protons are ranged out, e.g. with
20\,mm of copper absorber, 99.8\% (99.5\%) of the energy deposited in the
on-axis (off-axis) scintillator is prompt, i.e. within 1\,$\mu$s of the
beam pulse, making the details of neutron propagation and material activation
completely unimportant.

\subsubsection{Straggling}

To test if \geant's model for range straggling --- the variation in range for particles 
with same initial energy --- biases the result, the experiment was
first simulated with the extreme case of no straggling, such that the only significant
variations in protons' energy reaching the scintillator were from differing
path lengths due to multiple scattering and from the energy spread in the
initial beam.  Using these simulations to analyze the data instead of the
standard simulations resulted in an downwards shift of 4.7\,eV in the inferred
\ivalue. As a less extreme case, the \geant \texttt{UniversalFluctuations} model was substituted for
more accurate \texttt{UrbanFluctuations} model (which is the default for both
\texttt{Opt0} and \texttt{Opt4}/\texttt{EMZ}), which resulted in a downwards
shift of 0.5\,eV.  From this, we conclude that further improvements in the
straggling model are unlikely to result in shifts of more than half this amount
and adopt 0.3\,eV as the systematic error from the straggling model.

\subsubsection{Further numerical approximations}

Two further options that, in principle, increase \geant's accuracy were tested.  First, the
number of bins per energy decade for the dE/dx table was increased from its default of 20 to 100.
No significant change was seen in the output, and the effect on the \ivalue is less than 0.2\,eV.
Similarly, the production cuts were reduced from their default value of 0.7\,mm to 0.007\,mm.
Again no significant change was seen and the impact is less than 0.2\,eV.

\subsection{Alignment}\label{sec:align}

\subsubsection{Vessel alignment}

All experimental components were aligned to better than $1^\circ$ with the
beam. Any unmodeled misalignment would increase the protons' path length and
make the inferred I-value lower.  If the liquid argon vessel were misaligned by
$1^\circ$, it would lower the I-value by 0.3\,eV, while a similar misalignment 
in the copper strips would shift the inferred I-value by less than 0.1\,eV.

The vessel's vertical position was determined to 0.5\,cm.  Similarly to the case
of the liquid argon depth, above, there is some slight sensitivity to the vertical position
even if all straight-line proton paths lead entirely through argon because protons can
scatter from the argon into the aluminum vessel and then back into the argon.  As the
argon-aluminum boundary is much closer to the beam center than the argon-air boundary,
there is more possibility of this effect being significant.  Simulation
showed that the effect on the \ivalue is only $\pm 0.2$\,eV.

\subsubsection{Scintillator position}

The scintillator panels were placed with an accuracy of 1\,cm.  By analyzing the data
under the assumption that the scintillator is out of its nominal position, it was
determined that very little uncertainty is incurred, with the answer only shifting
by 0.2\,eV.

\section{\label{sec:conclusions} Conclusions}

The present results are shown in \fig{evidence} in comparison to the previous
evaluations in Ref.~\cite{self} for gaseous and liquid argon.  The overall
result from this measurement combines the on-axis and off-axis scintillator
results, assuming full correlation of systematic uncertainties, to arrive
at \larmeas.

\figureevidence

To make our overall recommendation for the \ivalue of liquid argon, we combine
our previous evaluation for liquid argon~\cite{self} with the present results.
No significant sources of correlated uncertainties were identified between the
present measurement and the previous evaluation from Ref.~\cite{self}, which
relied primarily on oscillator strength distributions (i.e. photoabsorption
cross sections) and an estimate of the phase effect using previous measurements
several other substances in gaseous and condensed forms.  Secondary
contributions to the previous evaluations come from consideration of periodic
trends, a Hartree-Fock wave function calculation, and a stopping power
measurement using natural alpha particles in gaseous argon.  Even smaller
contributions are from four other stopping power measurements.  Some
correlation could exist in the treatment of multiple Coulomb scattering in the
stopping power measurements.  However, this group of measurements collectively
make so little contribution that the result is unchanged to the last reported
decimal place if they are discounted entirely.  Therefore we treat the
evaluation and this measurement as independent.  To avoid loss of accuracy, we
carried all digits through the calculation, using \laranswertwodigits for the
previous evaluation and \larmeastwodigits for the present measurement.  The
combined recommendation is \recommendation.  All uncertainties are given at $1\sigma$ (68\% CL).

Using our recommended value rather than the old values of 188\,eV or 187\,eV
shifts energy reconstruction in liquid argon TPCs used for GeV-scale neutrinos
by around $(0.2\pm0.2)$\%, with the exact shift depending on the analysis method.  For instance,
the mass stopping power of a minimum-ionizing particle is reduced from 1.508\,MeV\,g$^{-1}$\,cm$^{2}$
to $(1.505\pm0.003)$\,MeV\,g$^{-1}$\,cm$^{2}$.  The
inferred value of $\Delta m^2_{32}$ is directly proportional to reconstructed
energy.  As DUNE projects a sensitivity to this parameter of around 0.5\%
(depending on exposure)~\cite{DUNE:2023nqi}, is it important that a reliable
value and uncertainty for the mean excitation energy of liquid argon be used.

\section{Acknowledgments}

This manuscript has been authored by Fermi Forward Discovery Group, LLC under
Contract No. 8924\,\allowbreak 3024\,\allowbreak CSC\,\allowbreak
000\,\allowbreak 002 with the U.S. Department of Energy, Office of Science,
Office of High Energy Physics.  Thanks to Chandrashekhara Bhat, Dali
Georgobiani, Carol Johnstone, Matthew King, Alajos Makovec, Lionel Prost,
Nathaniel Rowe,  Alexander Shemyakin, and Katsuya Yonehara for providing
invaluable information and assistance.

\bibliographystyle{JHEP}
\bibliography{larImeaspaper.bib}

\end{document}